\documentclass{aa}						% Required by A&A

\usepackage[english]{babel}   % English hyphenation etc.
\usepackage[utf8]{inputenc}   % Multi-byte encoding
\usepackage{amsmath,amssymb,amsfonts} % Better mathematics
\usepackage[version=3]{mhchem}           % Typeset chemical formulae easily e.g. \ce{H2SO4}
\usepackage{siunitx}					% Package for typesetting units correctly
\usepackage{multirow}         % Enables multirow and multicolumn spanning in tables
\usepackage[flushleft]{threeparttable}     % Allow for notes under a table
\usepackage{graphicx}         % Include external graphics files
\usepackage[varg]{txfonts}		% Required by A&A
\usepackage[usenames,dvipsnames,svgnames,table]{xcolor}

\usepackage{natbib}
\bibpunct{(}{)}{;}{a}{}{,}

\begin{document}

%%%%%%%%%%%%%%%%%%%%%%%%%% MANUSCRIPT HEADER %%%%%%%%%%%%%%%%%%%%%%%%%%%%%%%%%%%%%

\title{The slow spin of the young sub-stellar companion \\ GQ Lupi b and its orbital configuration}
%\subtitle{My Subtitle}
\author{Henriette Schwarz\inst{1}\thanks{Email: schwarz@strw.leidenuniv.nl}
\and Christian Ginski\inst{1}
\and Remco J. de Kok\inst{1,2}
\and Ignas A. G. Snellen\inst{1}
\and \\ Matteo Brogi\inst{3,5}
\and Jayne L. Birkby\inst{4,6}}
\institute{Leiden Observatory, Leiden University, PO Box 9513, 2300 RA Leiden, The Netherlands
\and SRON Netherlands Institute for Space Research, Sorbonnelaan 2, 3584 CA Utrecht, The Netherlands
\and Center for Astrophysics and Space Astronomy, University of Colorado at Boulder, CO 80309 Boulder, USA
\and Harvard-Smithsonian Center for Astrophysics, 60 Garden Street, MA 02138 Cambridge, USA
\and NASA Hubble Fellow
\and NASA Sagan Fellow }
\date{}

\abstract
{The spin of a planet or brown dwarf is related to the accretion process, and therefore studying spin can help promote our understanding of the formation of such objects. We present the projected rotational velocity of the young sub-stellar companion GQ Lupi b, along with its barycentric radial velocity. The directly imaged exoplanet or brown dwarf companion joins a small but growing ensemble of wide-orbit sub-stellar companions with a spin measurement. The GQ Lupi system was observed at high spectral resolution ($R \sim$\num{100000}), and in the analysis we made use of both spectral and spatial filtering to separate the signal of the companion from that of the host star. We detect both \ce{CO} (S/N=11.6) and \ce{H2O} (S/N=7.7) in the atmosphere of GQ Lupi b by cross-correlating with model spectra, and we find it to be a slow rotator with a projected rotational velocity of $5.3^{+0.9}_{-1.0}$ \si{\km\per\s}. The slow rotation is most likely due to its young age of $< \SI{5}{Myr}$, as it is still in the process of accreting material and angular momentum. We measure the barycentric radial velocity of GQ Lupi b to be \SI{2.0 \pm 0.4}{\km\per\s}, and discuss the allowed orbital configurations and their implications for formation scenarios for GQ Lupi b.}


\maketitle

%%%%%%%%%%%%%%%%%%%%%%%%%% MAIN TEXT %%%%%%%%%%%%%%%%%%%%%%%%%%%%%%%%%%%%%%%%%%%%%

\section{Introduction}

Measurements of the spin and the orbit of giant extrasolar planets and brown dwarf companions may hold important clues to their origin and evolution. Generally, two formation processes are considered for giant planets: i) core accretion and ii) disk fragmentation. Jupiter and Saturn are commonly accepted to have formed through core accretion, a class of formation models where gas accretes onto solid planetary embryos of several to ten Earth masses which may have formed beyond the iceline by runaway accretion from kilometer-sized planetesimals \citep{Pollack1996, Laughlin2004, Hubickyj2005}. The discovery of extrasolar giant planets led to the reinvigoration of the disk fragmentation hypothesis which states that giant (exo)planets may form as a disk gravitational instability that collapses on itself in the outer protoplanetary disk \citep{Boss1997, Boss2000}. Disk fragmentation is also considered a potential formation scenario for the more massive brown dwarf companions \citep{Chabrier2014}, or alternatively brown dwarf companions can be the result of prestellar core fragmentation during the earliest stages of the cloud collapse \citep{Jumper2013}.

Spin is predominantly a result of accretion of angular momentum during the formation, and if core accretion and gravitational instability result in differences in spin angular momentum, it is possible this will show up in studies of spin of sub-stellar companions as function of mass. In the Solar System, the spin angular momenta of those planets not influenced by tidal effects or tidal energy dissipation by a massive satellite follow a clear relationship, spinning faster with increasing mass \citep{Hughes2003}. In particular, gas giants far away from their central star are likely to have primordial spin angular momentum, making the directly imaged sub-stellar companions ideal candidates for exploring the connection between formation and spin.

The rotational velocity of an exoplanet was measured for the first time by \citet{Snellen2014}, observing the directly imaged planet $\beta$ Pictoris b with high-dispersion spectroscopy, measuring it to have a projected rotational velocity of $v\sin(i)=$ \SI{25}{\kilo\m\per\s}. Another young directly imaged planet, 2M1207 b, became the first exoplanet to directly have its rotational period measured ($P_{\textrm{rot}}=$ \SI{10.7}{hr}), when \citet{Zhou2016} detected rotational modulations in HST/WFC3 photometric monitoring of the object. Both results are in accordance with an extrapolation of the spin-mass trend observed in the Solar System planets (see Fig. \ref{fig:planet_spin}). 

Apart from being related to the accretion process, the spin of an exoplanet is also a fundamental observable that affects in particular its atmospheric dynamics and climate as well as e.g. its magnetic fields. On Earth, the Coriolis effect generates large-scale ocean currents which in turn promote cyclones. For fast rotators, including many brown dwarfs, the wind flows are rotation dominated \citep{Showman2013}. On the other hand, exoplanets orbiting close to their parent star are expected to be tidally locked. \citet{Brogi2016} and \citet{Louden2015} both recently made use of high-dispersion spectroscopy to detect a Doppler signature in the transmission spectrum of the hot Jupiter HD 189733 b, consistent with synchronous rotation. Synchronous rotation is the cause of large temperature differences between the day- and night-side which in turn can cause fast winds flowing from the hot day-side to the cold night-side.

Another approach to understanding the formation and dynamical evolution of exoplanets is to study their orbits. In the case of directly imaged sub-stellar companions, it is often difficult to constrain the orbits, due to the long time-scales involved \citep[e.g.][]{Pearce2015, Ginski2014A}. However, using a high-dispersion slit spectrograph in combination with adaptive optics, it is possible to extract spatially resolved high-dispersion spectra for the companion \citep{Snellen2014}, and thereby measure even very small Doppler shifts due to the orbital motion of the companion. Thus, even just one radial velocity measurement can in some cases prove to be a powerful orbital constraint \citep{Nielsen2014, Lecavelier2016}.

In this paper we present both the spin measurement and the barycentric radial velocity of the widely-separated sub-stellar companion GQ Lupi b from high-dispersion spectroscopy. We introduce the GQ Lupi system in Section \ref{sec:The-GQ-Lupi-system}, and give the details of the observations in Section \ref{sec:Observations}. The data analysis is detailed in Section \ref{sec:Data-analysis} with special emphasis on how the spatially resolved spectrum of the companion is extracted and cleaned from telluric and stellar spectral lines. In Section \ref{sec:Measuring-the-signal} we explain the cross-correlation analysis which is employed to measure the rotational broadening and doppler shifts of the molecular lines in the companion spectrum, and the results from the companion are presented in Section \ref{sec:Results}, along with the host star spin and systemic velocity. We discuss the implications of the spin measurement in Section \ref{subsec:The-slow-spin} and the constraints on orbital elements in Section \ref{subsec:The-orbital-orientation}.

\section{The GQ Lupi system}
\label{sec:The-GQ-Lupi-system}

\object{GQ Lupi} A is a classical T Tauri star with spectral type K7\,V \citep{Kharchenko2009}, located in the star forming cloud Lupus I \citep{Tachihara1996} at an approximate distance of \SI{140 \pm 50}{\pc} \citep[e.g.][]{Neuhauser1998}. The system is less than five million years old \citep{Neuhauser2005, SeperueloDuarte2008, Weise2010}, and there is strong observational evidence for a circumstellar warm dust disk \citep{Hughes1994, Kessler-Silacci2006, Morales2012, Donati2012}, extending to between \num{25} and \SI{75}{\au} from the star \citep{Dai2010}. The inclination of the spin axis of GQ Lupi A is estimated to be \ang{30} \citep{Donati2012}, which is roughly consistent with the inclination of the warm inner parts of the circumstellar disk determined by \citet{Hugelmeyer2009} to be $\sim$\ang{22}. Additional stellar parameters are given in Table \ref{table:GQLupiA}.

The star has a directly imaged sub-stellar companion, GQ Lupi b, with a highly uncertain and model dependent mass most likely in the range \SIrange{10}{36}{\Mjup}, \citep{Marois2007, Seifahrt2007}, placing it somewhere on the border between a giant planet and a brown dwarf, although this boundary is becoming increasingly blurred \citep{Chabrier2014, Hatzes2015}. The exoplanet candidate was first discovered with HST by \citet{Neuhauser2005}, and it was recently imaged as part of the SEEDS survey \citep{Uyama2016}. It is located at a projected separation of \ang{;;0.7} west (\SI{100}{\au} at \SI{140}{\pc}) and an astrometric analysis by \citet{Ginski2014B} points to a best fit semi-major axes of \SIrange{76}{129}{\au} (at \SI{140}{\pc}) and high eccentricity in the range \numrange{0.21}{0.69}. The GQ Lupi system has a favourable companion to star contrast ratio, with the host star \citep{Kharchenko2009} and companion \citep{Ginski2014B} having K-band magnitudes of \num{7.1} and \num{13.3}, respectively. Selected parameters for GQ Lupi b are given in Table \ref{table:spin}.

\begin{table}[ht]
\caption{Parameters for GQ Lupi A from \citet{Donati2012}}  	
\label{table:GQLupiA}      								
\centering       
\begin{tabular}{l r c l}          	
\hline\hline                            
\\
Photospheric temerature, $T_{\textrm{eff}}$ & $4300$   & $\pm$ & \SI{50}{\K}         \\
Stellar mass, $M_{\star}$                   & $1.05$   & $\pm$ & \SI{0.07}{\Msun}    \\
Stellar radius, $R_{\star}$                 & $1.7$    & $\pm$ & \SI{0.2}{\Rsun}     \\
Surface gravity, $\log g$                   & $3.7$    & $\pm$ & \num{0.2}               \\    
Rotation period, $P_{\star}$                & $8.4$    & $\pm$ & \SI{0.3}{days} \\
Proj. rot. velocity, $v\sin(i)$    & $5.0$    & $\pm$ & \SI{1.0}{\km\per\s} \\
Inclination, $i$                            & \ang{30} &       &  \\
\hline                            
\end{tabular}
\end{table}

\section{Observations}
\label{sec:Observations}

We observed the GQ Lupi system for 1 hour (including acquisition) on 29 May 2014 with the Cryogenic High-Resolution Infrared Echelle Spectrograph \citep[CRIRES,][]{Kaeufl2004}, located at the Nasmyth A focus of VLT Antu of the European Southern Observatory at Cerro Paranal in Chile. CRIRES is a long slit spectrograph with a slit length of \ang{;;50}, and we chose a slit width of \ang{;;0.2} to achieve the maximal spectral resolving power $R \sim$\num{100000}. The slit was positioned in a near east to west orientation to both encompass the host star and the substellar companion located at position angle PA = \ang{-83.6} \citep{Zhou2014}, allowing a combination of high dispersion spectroscopy with high-contrast imaging \citep{Snellen2015}. 

We took 18 exposures of \SI{120}{\s} in a classical ABBA pattern, where the telescope was nodded sequentially \ang{;;10} in the slit direction between postions A and B, then B and A to allow accurate background subtraction. At each postion an additional small random jitter offset was introduced to minimise issues from flatfielding and hot pixels.

The four CRIRES detectors are of type Aladdin III InSb, and each has a size of \num{1024 x 512} pixels with a gap of approximately \num{280} pixels in between. Unfortunately, the two outer detectors have severe odd-even column non-linearity effects\footnote{\url{http://www.eso.org/sci/facilities/paranal/
instruments/crires/doc/VLT-MAN-ESO-14500-3486_v93.pdf}} which were not possible to accurately calibrate in these data. These are therefore left out from the analysis. The observations were performed with the standard wavelength settings for order 24 ($\lambda_{\textrm{cen}} = \SI{2.3252}{\micro\m}$), using the two central detectors to target the ro-vibrational (2, 0) R-branch of carbon monoxide. The wavelength range \SIrange[range-phrase = --]{2.302}{2.331}{\micro\m} of the two central detectors covers more than 30 strong \ce{CO} lines.  

We used the Multi Application Curvature Adaptive Optics system \citep[MACAO,][]{Arsenault2003} in \ang{;;1.0} to \ang{;;1.1} seeing conditions. To further maximize the performance of MACAO, the target system was observed at low airmasses between 1.056 and 1.121. This resulted in the starlight being suppressed by a factor $\sim 45$ at the companion position. With a K-band contrast ratio of \num{3.3e-3}, the companion contributes with $(F_b/F_A)*45*100 =\SI{15 \pm 3}{\%}$ of the total flux at the companion position. The remaining $85\%$ flux is from the host star.

\begin{figure}[ht]
\resizebox{\hsize}{!}{\includegraphics{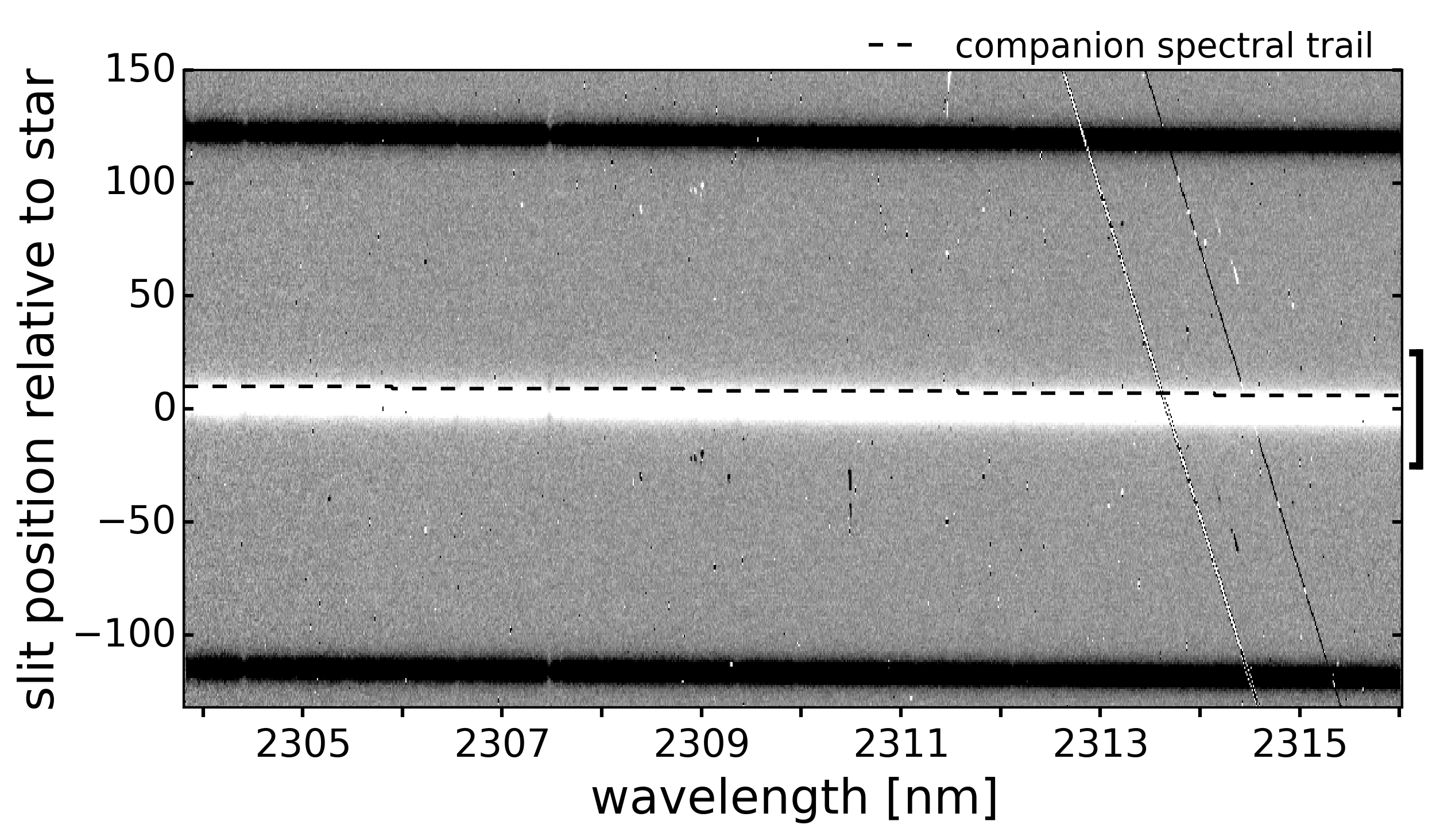}}
	\caption{One of the combined image frames from detector 2. This is an intermediate data product from the CRIRES pipeline, and it is the combination of an AB nodding pair of exposures. The central rows that are used in the further data analysis are indicated on the right axis. The stellar spectral trail is clearly visible as the central white band, and the location of the hidden spectral trail from GQ Lupi b is indicated with a dashed line. The two diagonal stripes towards the right end are detector defects.}  
	\label{fig:comb_illustration}
\end{figure}

\section{Data analysis}
\label{sec:Data-analysis}

\subsection{Basic data reduction}
\label{subsec:Basic-data-reduction}

The data was processed with the CRIRES pipeline version 2.3.2 and the corresponding version 3.10.2 of ESOREX. The pipeline performed the basic image processing, i.e. the images were dark subtracted, flatfielded and were corrected for known bad pixels and non-linearity effects. Furthermore, the pipeline combined the images in AB nodding pairs performing a background subtraction, and extracted a 1D spectrum from each of these combined images with the optimal extraction technique \citep{Horne1986}. 

We made use of the intermediate data products i.e. the combined images, as well as the optimally extracted 1D spectra of the host star as follows. From the 18 exposures, 9 combined image frames were produced for each detector. An example of such a frame for detector 2 is shown in Fig. \ref{fig:comb_illustration}. We cut away everything but the central 51 rows that contain the stellar spectral trail, the hidden companion spectral trail (8 pixels above the center), and enough extra rows to properly determine the stellar point spread function along the slit. The bad pixel correction in the CRIRES pipeline is insufficient. Therefore remaining bad pixels (including cosmic rays) were visually identified with the program DS9, and they were corrected with cubic spline interpolation using the four nearest neighbours in the row on both sides.

The 9 optimally extracted 1D spectra of the host star were corrected for bad pixels in the same manner, and then median normalised. Subsequently they were averaged over time to a single reference spectrum, representing the average spectrum of the host star, GQ Lupi A, plus telluric absorption in the Earth's atmosphere.

The wavelength calibration was performed using line matching between deep and isolated telluric lines in the reference spectrum and a synthetic transmission spectrum from ESO SkyCalc\footnote{https://www.eso.org/observing/etc/skycalc/}. We used 15 to 20 lines per detector and fitted a second-order polynomial to the pairs of pixel and wavelength centroids, where the centroids were determined from Gaussian fits. The highest residuals to the second-order polynomial fits were at a level of $20\%$ of a pixel.

\begin{figure}[ht]
\resizebox{\hsize}{!}{\includegraphics{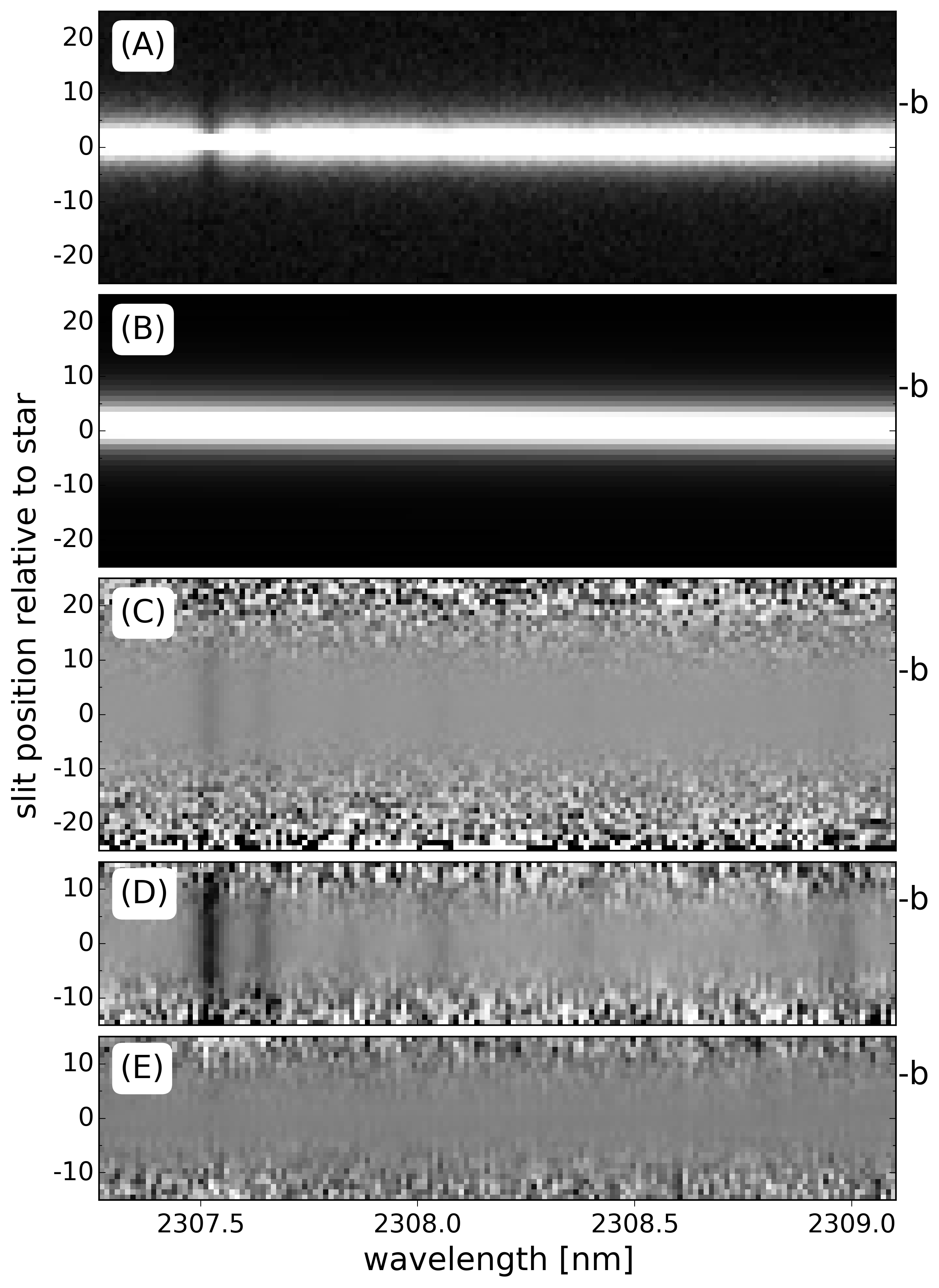}}
	\caption{Illustration of the data analysis steps from a short wavelength range from  detector 2. The y-axis depicts the slit position relative to the star, with the slit position of the companion indicated by a 'b' on the right-hand axes. The grey-scales have been adjusted individually for each array with an IRAF-like z-scale algorithm. Top to bottom: (A) The AB-combined image corrected for bad pixels, (B) The spatial profiles of the star, (C) The image after each row has been normalised with the rows of the spatial profile array, (D) The spatial spectra (i.e. rectified for the trace), (E) The residual spatial spectra after removing stellar and telluric lines.}  
	\label{fig:waterfall}
\end{figure}

\subsection{Extraction of spectra for each slit position}
\label{subsec:Extraction-of-spectra}

From the clean combined image frames (Fig. \ref{fig:waterfall}A), we optimally extracted a spectrum for each slit position. Fig. \ref{fig:waterfall} illustrates the procedure, which was performed individually for each frame and each detector. Looking at the star trail in Fig. \ref{fig:comb_illustration}, it can be seen that the y-position of the intensity peak shifts with wavelength, and it is an important step in the data analysis to correct for this tilt of the trail of the star. In detector 2, the pixel with the maximum intensity shifts by 4 pixels from one end of the detector to the other, and in detector 3 the shift is 2 pixels. The tilted trail introduces a curvature in the continuum for a given row, and therefore we normalised each row with a polynomial fit of the curvature, where the degree of the polynomium depended on the distance from the stellar spectral trail. The fitted curves also provide the spatial profiles of the star as a function of wavelength (Fig. \ref{fig:waterfall}B). We used the spatial profiles to optimally extract a spectrum for each of the 31 most central rows from the normalised arrays (Fig. \ref{fig:waterfall}C). We will refer to these collectively as the spatial spectra (Fig. \ref{fig:waterfall}D). The 9 AB frames of spatial spectra were combined to a single frame to increase the signal-to-noise ratio of a given spectrum. Each pixel was normalised by the median of the pixel value through the 9 frames and then combined as a weighted average. The weights were determined according to the varying width of the spatial profile (Fig. \ref{fig:waterfall}B) which is a seeing proxy. Pixels that deviated $>4\sigma$ for the given pixel position were excluded.

\subsection{Removal of telluric and stellar spectrum}
\label{subsec:Removal-of-telluric-and-stellar-spectrum}

The spatial spectra are dominated by the telluric absorption and the stellar spectrum which has prominent \ce{CO} lines. In contrast, the additional component to the spectra from the companion is strongly localised on the detector at the slit position of the companion. The stellar and telluric components are quasi identical at all slit positions, although the spectral resolution changes slightly with slit position. This is due to variations in the spatial profile of the star along the slit. Furthermore a small offset of the position of the star with respect to the center of the slit can result in a small wavelength offset, and in addition there is a scaling factor. In order to remove both the telluric and stellar components, we made use of the reference spectrum we constructed in Section \ref{subsec:Basic-data-reduction}. The reference spectrum was adjusted to each slit position to correct for the above mentioned effects by convolving with an appropriate broadening function, which was determined with the singular value decomposition technique \citep{Rucinski1999}. The telluric and stellar spectrum was then removed from the spatial spectra by dividing with the adjusted reference spectrum (Fig. \ref{fig:waterfall}E).

\subsubsection{Companion position}
\label{subsec:Companion-position}

The presence of \ce{CO} lines in the spectrum of the host star is a complicating factor, because they potentially overlap with the \ce{CO} lines of the companion. If the orbital motion of the companion has a significant radial component, the lines will be red- or blueshifted relative to the lines of the star, but because of the wide separation between host and companion in the GQ Lupi system, the upper limit on the absolute value of this shift is \num{2.5} to \SI{3.5}{\km\per\s}, depending on the exact semi-major axis and mass of GQ Lupi A. Furthermore, both the molecular lines of the companion and of the host star are rotationally broadened. \citet{Guenther2005} measured the projected rotational velocity of GQ Lupi A to be \SI{6.8 \pm 0.4}{\km\per\s}, and more recently 
\citet{Donati2012} found a similar value of \SI{5 \pm 1}{\km\per\s}. The concern is that the shape and strength of the companion lines may be affected by the process of removing the stellar lines, thereby compromising the $v\sin(i)$ and/or Doppler shift measurements.

The severity of this issue is greatly reduced by the angular separation between the host and the companion of \ang{;;0.7}. As stated in section \ref{sec:Observations}, we achieved a suppression of the starlight at the position of the companion to a few percent, which means that $15\%$ of the total flux at this slit position originates from the companion and $85\%$ originates from the host star. We therefore scaled the stellar lines in the reference spectrum down by $15\%$ before proceeding as described in section \ref{subsec:Removal-of-telluric-and-stellar-spectrum} with removing the stellar and telluric spectrum from the spatial spectrum of the companion position.

In order to rescale the stellar reference spectrum, we isolated the stellar lines from the telluric parts of the reference spectrum. This was done as an iterative process, where a telluric model spectrum and a stellar model spectrum were fitted and removed separately from the reference spectrum. The steps were as follows:

\begin{itemize}\itemsep1.2pt
 \item We used the same telluric model spectrum from ESO SkyCalc which was used to perform the wavelength calibration. The airmass and the precipitable water vapor (PWV) was set manually at airmass 1.1 and PWV 1.5 to best fit the observed reference spectrum. The stellar model is a PHOENIX spectrum from \citet{Husser2013} with $T_{\textrm{eff}}=\SI{4300}{\K}$ and $\log g=\num{3.5}$. The telluric model was smoothed, and both the telluric and the stellar model were resampled to the observed wavelength solution.
 
 \item We convolved the telluric model with a broadening kernel to adjust the resolution to that of the reference spectrum. As in section \ref{subsec:Removal-of-telluric-and-stellar-spectrum} the singular value decomposition (SVD) technique \citep{Rucinski1999} was employed to determine the appropriate kernel, but here it was done in log space to avoid the deepest telluric lines from dominating. The telluric spectrum was then removed from the reference spectrum by dividing through with the telluric fit. 
 
 \item The stellar template was fitted to the telluric-removed reference spectrum by convolving with an SVD broadening kernel in log space. Following this, the stellar lines were removed from the original reference spectrum by dividing through with the stellar fit. 
 
 \item The telluric SVD fit and removal was redone, adjusting the telluric model to the stellar-removed reference spectrum before dividing the original reference spectrum through with the fit, thus producing an improved telluric-removed reference spectrum.
 
 \item The continuum of the new telluric-removed reference spectrum was fitted with a second order polynomial and removed. Subsequently the stellar SVD fit was redone, fitting the stellar model to the flat telluric-removed reference spectrum. This stellar fit was then scaled to $15\%$ of the line strength, and the original reference spectrum was divided through with this scaled down stellar model, resulting in a reference spectrum with the strength of the stellar lines reduced by $15\%$. 

\end{itemize}

The effects of scaling down the stellar lines are minor. Details of how it affects the measured $v\sin(i)$ and $RV$ are given in Section \ref{subsec:The-slow-spin} and Section \ref{subsec:The-orbital-orientation}.

\begin{figure}[ht]
\resizebox{\hsize}{!}{\includegraphics{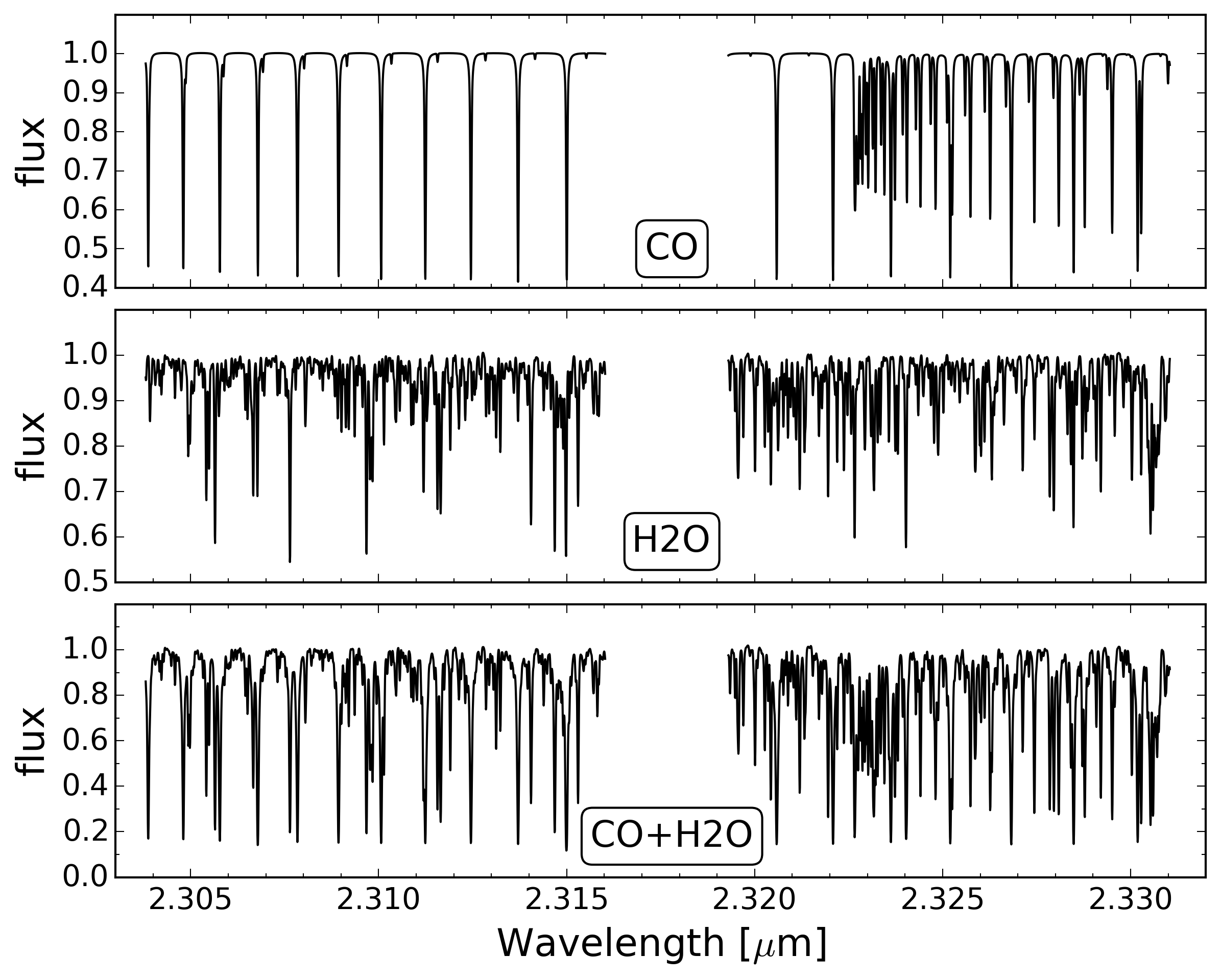}}
	\caption{Models with a T/p profile and abundance that give rise to the strongest cross-correlation signals from the companion. The top panel is the model with \ce{CO} as a single trace gas, the middle panel the one with only \ce{H2O}, and the bottom panel model contains both \ce{CO} and \ce{H2O}. The temperature is set to decrease from \SI{2150}{K} at \SI{1}{bar} to \SI{1100}{K} at \SI{0.03}{bar} and is isothermal outside this pressure range. The volume mixing ratios for \ce{CO} and/or \ce{H2O} in these three models are \num{e-4}.}
	\label{fig:3models_illustration}
\end{figure}

\begin{figure*}[ht]
\centering
\includegraphics[width=17cm]{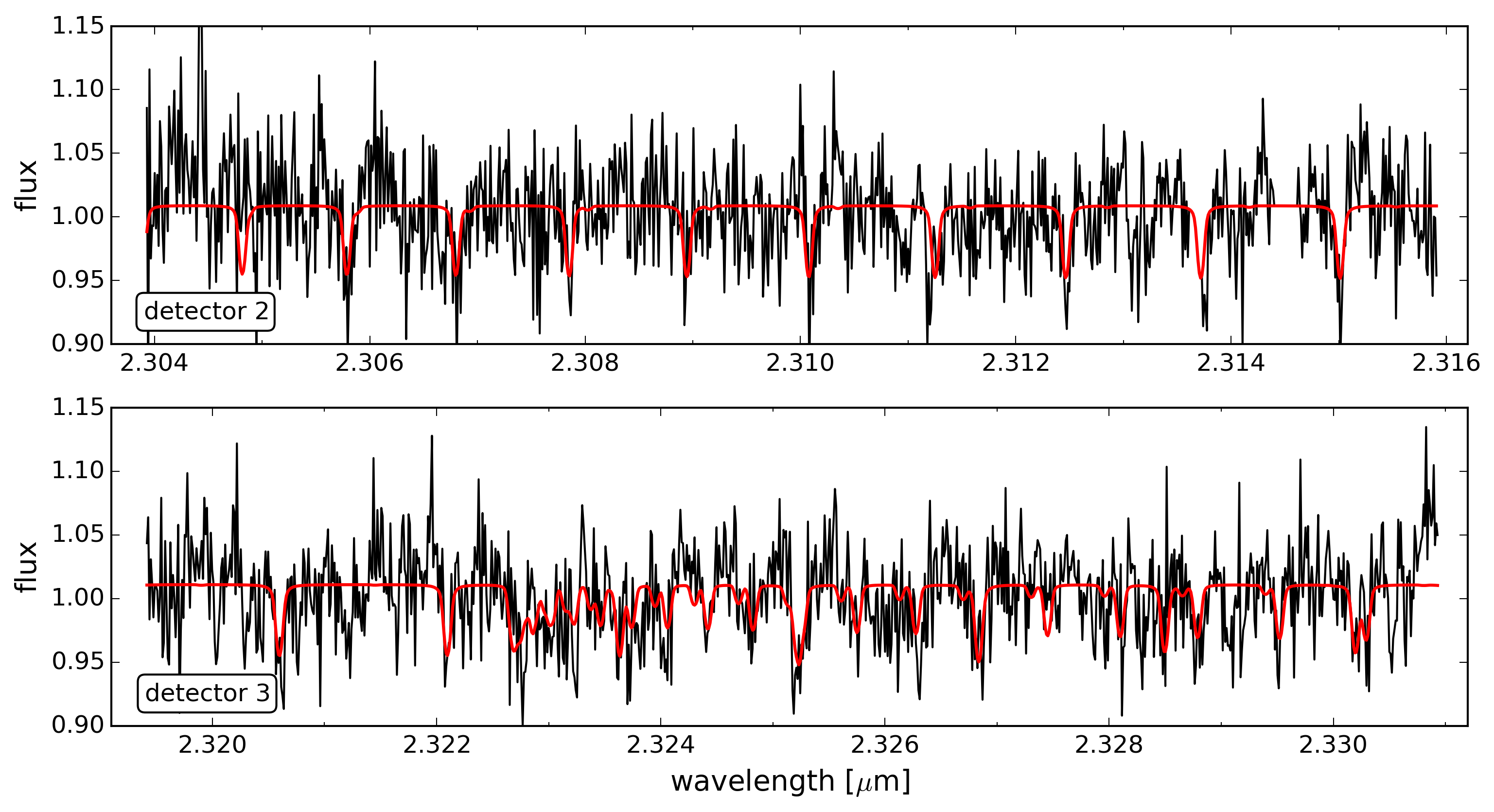}
	\caption{The residual spectrum after removing telluric and stellar lines at the companion position. Overplotted in red is the CO model from the top panel of Fig. \ref{fig:3models_illustration}, convolved to the CRIRES resolution, rotationally broadened, and Doppler shifted to match the measured $v\sin i$ (\SI{5.3}{\km\per\s}) and radial velocity (\SI{2}{\km\per\s}) of GQ Lupi b. For illustration purposes the model was also fitted to the residual companion spectrum with a vertical offset and a scaling factor. The companion spectrum is dominated by noise, but the individual CO lines are discernible.}  
	\label{fig:companion-spectrum}
\end{figure*}

\section{Measuring the signal from the companion}
\label{sec:Measuring-the-signal}

After removing the telluric and stellar spectrum, the residual spatial spectra consist of residual noise, with the exception at the slit position of the companion which also has a contribution from the companion spectrum. Fig. \ref{fig:waterfall}E shows a short wavelength range of the residual spatial spectra. At the high spectral resolution of the CRIRES data, the molecular bands are resolved into individual lines, and the signal from all of the lines from a given molecule within the wavelength range can be combined through cross-correlation with a model spectrum. Molecules are identified in the spectrum of the companion as a peak in the cross-correlation function (CCF). The CCF profile is sensitive to both the shape and the Doppler shift of the companion spectral lines. Lines that are broadened by rotation will result in broadened CCF profiles, and lines that are shifted will also shift the peak of the CCF. In this section we describe the model spectra and the cross-correlation analysis, as well as the procedure for measuring the rotation and radial velocity of the companion from the cross-correlation function.

\subsection{The model spectra}
\label{subsec:The-model-spectra}

The GQ Lupi system is thought to be very young ($<$\SI{5}{Myr}), and as a result GQ Lupi b is very hot with an effective temperature of approximately \SI{2650}{K} \citep{Seifahrt2007}. At such high temperatures the most abundant trace gas molecules in the K-band are expected to be \ce{CO} and \ce{H2O}. We have cross-correlated with models with \ce{CO} or \ce{H2O} as a single trace gas, as well as a suit of models containing both \ce{CO} and \ce{H2O}. The models were calculated line by line, assuming \ce{H2}-\ce{H2} collision-induced absorption \citep{Borysow2001, Borysow2002}. The \ce{CO} and \ce{H2O} data were taken from HITEMP 2010 \citep{Rothman2010}, and a Voigt line profile was employed. 

The models were constructed from a narrow grid of parameterised temperature-pressure profiles. All the models are isothermal with a temperature of $T_0$ = [\SI{1650}{K}, \SI{1900}{K}, \SI{2150}{K}] at pressures higher than $p_0$ = \SI{1}{bar}. The temperature decreases with a constant lapse rate (i.e. the rate of temperature change with log pressure) until it reaches $T_1$ = [\SI{750}{K}, \SI{1100}{K}, \SI{1450}{K}, \SI{1800}{K}] at pressure $p_1$ = [\SI{e-1.5}{bar}, \SI{e-2.5}{bar}, \SI{e-3.5}{bar}, \SI{e-4.5}{bar}]. At higher altitudes or equivalently lower pressures the models are again isothermal. Both \ce{CO} and \ce{H2O} were tested with four different volume mixing ratios, VMR = [\num{e-5.5}, \num{e-5.0}, \num{e-4.5}, \num{e-4.0}]. Fig. \ref{fig:3models_illustration} illustrates the \ce{CO}-only, \ce{H2O}-only and \ce{CO}+\ce{H2O} models with the best temperature-pressure profile (T/p) and volume mixing ratios from the grid.

\subsection{Cross-correlation analysis}
\label{Cross-correlation-analysis}

The model spectrum was first convolved to the CRIRES spectral resolution using a Gaussian filter, then Doppler-shifted over the range \num{-250} to \SI{250}{\kilo\m\per\s} in steps of \SI{1.5}{\kilo\m\per\s}, and for each velocity step cross-correlated with the residual spatial spectra. We have assumed R = \num{100000} for the instrumental profile, but we have measured the actual resolving power to be in the range \numrange{80000}{90000}. This causes a small overestimation of $v\sin(i)$ on the order of \SI{60}{\m\per\s}, which is included in the lower uncertainty bound of the final result. The cross-correlation was performed for every slit position as a means to investigate the strength of spurious signals and issues with e.g. stellar residual signals. Each detector was treated separately, and only combined as an average after the cross-correlation analysis. This results in two-dimensional arrays with the cross-correlation coefficient as a function of slit position and applied Doppler-shift.

\begin{figure}[ht]
\resizebox{\hsize}{!}{\includegraphics{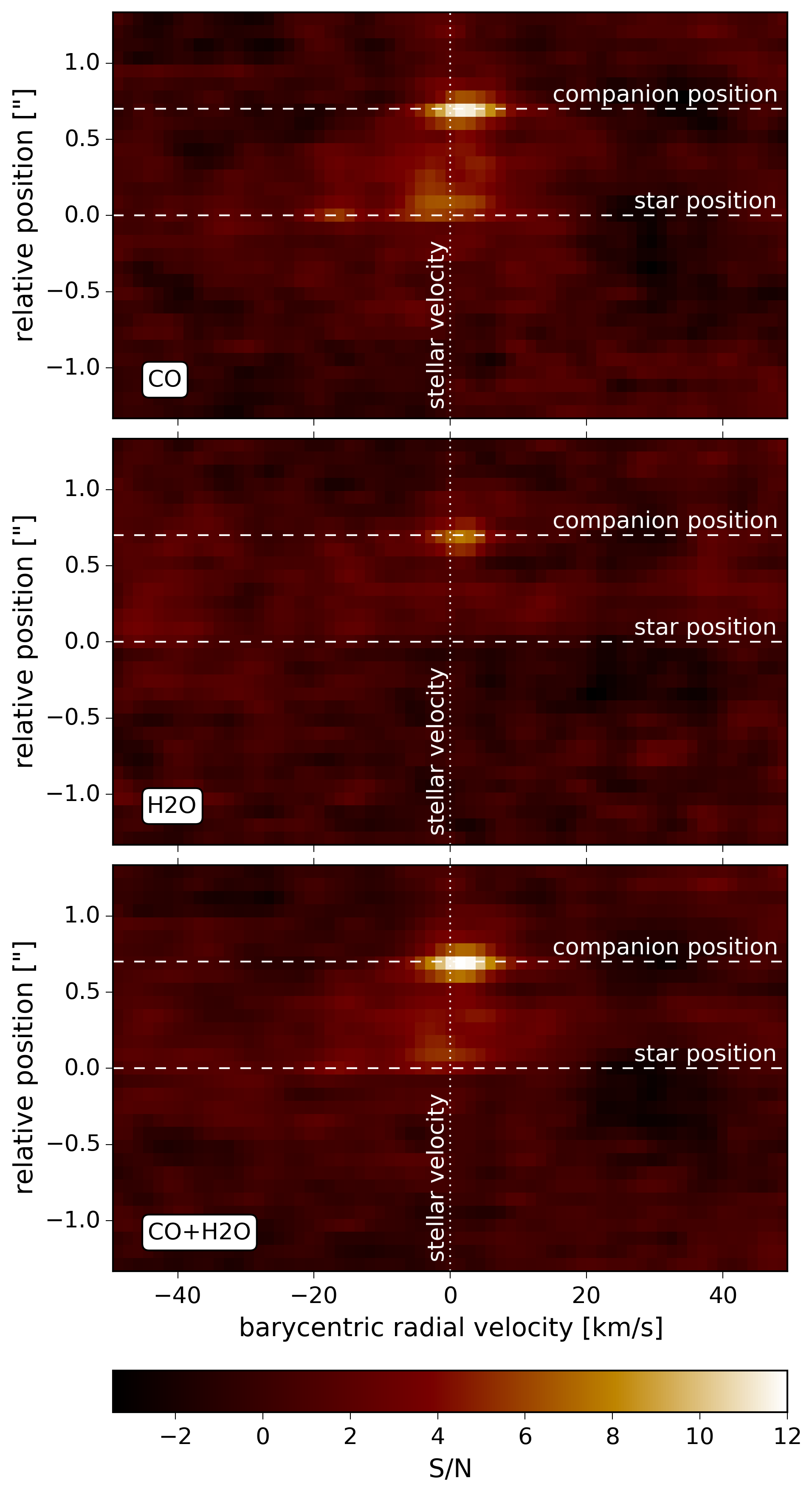}}
	\caption{Strength of the cross-correlation as function of slit position and radial velocity for the three best-fit models (see Fig. \ref{fig:3models_illustration}). We detect both \ce{CO} and \ce{H2O} using single-trace gas models with a S/N of 11.6 for \ce{CO} and \ce{7.7} for \ce{H2O}, and the signal from the double-trace gas model is marginally stronger with a S/N of 12.3. These cross-correlation arrays are the averages from the two central CRIRES detectors.}
	\label{fig:colormap}
\end{figure}

\subsection{Measuring the companion $v\sin(i)$ and RV}
\label{subsec:Measuring-the-companion-parameters}

The cross-correlation function (CCF) from the residual spectrum at the companion position using any of the models contains the rotational broadening and radial velocity of the companion. For the purpose of measuring $v\sin(i)$ and RV we selected the model with the strongest cross-correlation signal. The CCFs from the two detectors were averaged to further maximise the signal-to-noise (S/N). We will refer to the average CCF from the companion position and the best model as the measured companion CCF.

We determined the best fit $v\sin(i)$ and RV of the measured companion CCF through $\mathcal{X}^2$ minimisation with a suite of model CCFs, along with confidence intervals from rescaling the errors so $\bar{\mathcal{X}}^2 = 1$. The model CCFs were constructed by cross-correlating the non-broadened best model with broadened and shifted versions of itself. Both the broadened and non-broadened models were convolved to the CRIRES spectral resolution prior to the cross-correlation. We tested projected rotational velocities in the range \num{0} to \SI{10}{\km\per\s} and Doppler-shifts in the range \num{-5} to \SI{5}{\km\per\s}. For both parameters the stepsize was \SI{0.1}{\km\per\s}. Each of these model CCFs were then offset (y-direction) and scaled with a least square fit to best match the measured companion CCF, after which the $\mathcal{X}^2$ minimisation routine was performed. The measured radial velocity was corrected to the barycentric radial velocity using the systemic velocity ($v_{\textrm{sys}}$) determined from from the host star spectrum as described in Section \ref{subsec:method-host-star-vsini-and-vsys}, and the heliocentric correction term for the time of observation. The results are presented in Section \ref{subsec:companion-vsini-and-rv} and discussed in Sections \ref{subsec:The-slow-spin} and \ref{subsec:The-orbital-orientation}.

\subsection{Measuring the systemic velocity and the host star $v\sin(i)$}
\label{subsec:method-host-star-vsini-and-vsys}

We took the same approach to measuring $v\sin(i)$ and RV of the host star as we did for the companion. In this case, the measured host star CCF was the cross-correlation function of the flat telluric-removed reference spectrum from the final step of Section \ref{subsec:Companion-position} and a synthetic PHOENIX spectrum from \citet{Husser2013} with $T_{\textrm{eff}}=\SI{4300}{\K}$ and $\log g= 3.5$. A suite of model CCFs were constructed from the PHOENIX model with a range of $v\sin(i)$- and RV-values, and we performed the $\mathcal{X}^2$ analysis for the host star, comparing the measured and modelled CCFs. The measured radial velocity was corrected to the barycentric radial velocity (i.e. the systemic velocity) using the heliocentric correction term for the time of observation. We carried out this analysis for PHOENIX models with $T_{\textrm{eff}} \pm \SI{100}{\K}$ and $\log g \pm 0.5$ to test the sensitivity to the stellar parameters, and found that this temperature difference can affect the radial velocity with up to \SI{100}{\m\per\s} and the $v\sin(i)$ with up to \SI{200}{\m\per\s}. The radial velocity is not sensitive to the choice of $\log g$, but the $v\sin(i)$ can be affected by up to \SI{200}{\m\per\s}. The results are presented and discussed in Sections \ref{subsec:result-host-star-vsini-and-vsys} and \ref{subsec:discussion-host-star-vsini-and-vsys}, respectively.

\section{Results}
\label{sec:Results}

Although the residual spectrum at the companion position is noisy, the absorption lines from \ce{CO} in the companion atmosphere are in some cases visible (Fig. \ref{fig:companion-spectrum}). To help guide the eye, we have overplotted the best \ce{CO} model shifted to the best fit radial velocity. Although we can visually detect the individual \ce{CO} lines, the companion spectrum is too noisy to allow the rotational broadening or the Doppler shift to be measured directly from the \ce{CO} lines, and we therefore need to make use of the CCF.

\subsection{Detection of \ce{CO} and \ce{H2O}}
\label{subsec:Detection}

We clearly detect the sub-stellar companion at the expected distance from the host star (\ang{;;0.7}) in the CCF arrays (see Fig. \ref{fig:colormap}). We detect \ce{CO} with a S/N of 11.6 and \ce{H2O} with a S/N of 7.7. The best model containing both \ce{CO} and \ce{H2O} is detected with a S/N of 12.3. These are average values from the two central CRIRES detectors, but both \ce{CO} and \ce{H2O} are detected separately in each detector. The S/N-values were obtained by dividing the cross-correlation coefficients with the standard deviation of the array, excluding points in parameter space close to the companion signal. Fig. \ref{fig:colormap} shows the S/N cross-correlation arrays for the best \ce{CO}-only, \ce{H2O}-only and \ce{CO}+\ce{H2O} models from the tested T/p and VMR grid. They are displayed here from \num{-50} to \SI{50}{\km\per\s}, but we note that the S/N-values are based on the standard deviation of the full radial velocity range from \num{-250} to \SI{250}{\km\per\s}. These best models are illustrated in Fig. \ref{fig:3models_illustration} and share the same T/p profile with the temperature decreasing from \SI{2150}{K} at \SI{1}{bar} to \SI{1100}{K} at \SI{0.03}{bar}. The steep decrease in temperature relatively deep in the atmosphere gives rise to strong absorption lines. The VMR for both \ce{CO} and \ce{H2O} in these models are in all cases \num{e-4}. We note that all the models in the grid described in Section \ref{subsec:The-model-spectra} give rise to significant molecular detections, and for a given VMR the significance of a detection from the different T/p profiles agree to within $1\sigma$.

The \ce{CO} signal and the \ce{CO}+\ce{H2O} signal are detected with a stronger significance in detector 3 compared to detector 2. This is in line with expectations, because the third detector has a higher number of \ce{CO} lines. In detector 3 there are stellar \ce{CO} residuals in the row just above the star position, although at a lower level than the companion signal. The residuals are strong enough to also show up in the average cross-correlation arrays in Fig. \ref{fig:colormap}. It is unclear why the procedure for removing the stellar light has partly failed at this one slit position, and also why the issue only involves one of the detectors. The companion is a further 7 pixels away where the light from the star is decreased to only $\sim 2 \%$ of its peak intensity, so we expect this issue to have negligible effect on the $v\sin(i)$ and RV measurements.

\subsection{Companion $v\sin(i)$ and RV}
\label{subsec:companion-vsini-and-rv}

\begin{figure}[ht]
\resizebox{\hsize}{!}{\includegraphics{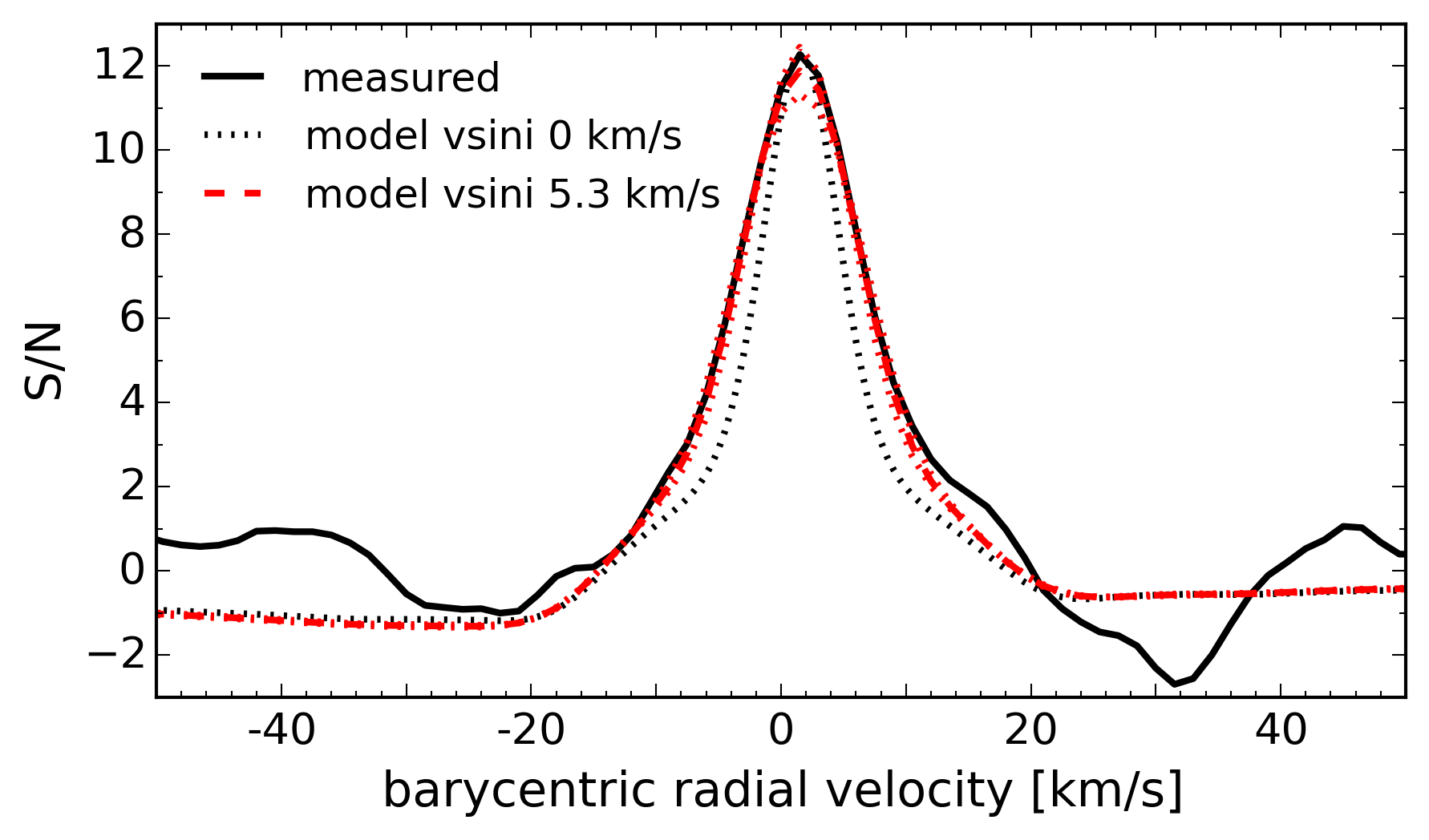}}
	\caption{The measured companion cross-correlation function (CCF) is best fit by a \ce{CO} + \ce{H2O} model CCF which is redshifted \SI{2.0}{\km\per\s} and rotationally broadened by \SI{5.3}{\km\per\s}. For comparison we also show the model CCF without rotational broadening for which the HWHM is dominated by the spectral resolution of the CRIRES instrument.}
	\label{fig:1D_best}
\end{figure}

Already from the Fig. \ref{fig:colormap} cross-correlation arrays it is clear that the companion signal is quite narrow, and that it is redshifted relative to the stellar radial velocity by only a small amount. The measured companion CCF is displayed in Fig. \ref{fig:1D_best} together with the best-fit model CCF, and the CCF for the same model, but without rotational broadening. We find from the $\mathcal{X}^2$ minimisation routine that the projected rotational velocity of GQ Lupi b is $5.3^{+0.9}_{-1.0}$ \si{\km\per\s}, and its barycentric radial velocity is \SI{2.0 \pm 0.4}{\km\per\s}.

\subsection{Host star $v\sin(i)$ and $v_{\textrm{sys}}$}
\label{subsec:result-host-star-vsini-and-vsys}

As a by-product of the analysis carried out in this work, we have estimates of the systemic velocity and the projected rotational velocity of GQ Lupi A. 
The systemic velocity is of direct importance as it is required to translate the measured companion RV to the barycentric frame. We find $v_\textrm{sys}=\SI{-2.8 \pm 0.2}{\km\per\s}$. For the host star, we have measured a rotational broadening corresponding to $v\sin(i)=\SI{6.8 \pm 0.5}{\km\per\s}$. However, we note that we have not taken additional broadening terms (e.g. macro-turbulence) into account when determining $v\sin(i)$ of GQ Lupi A.

\begin{figure*}[ht]
\centering
\includegraphics[width=17cm]{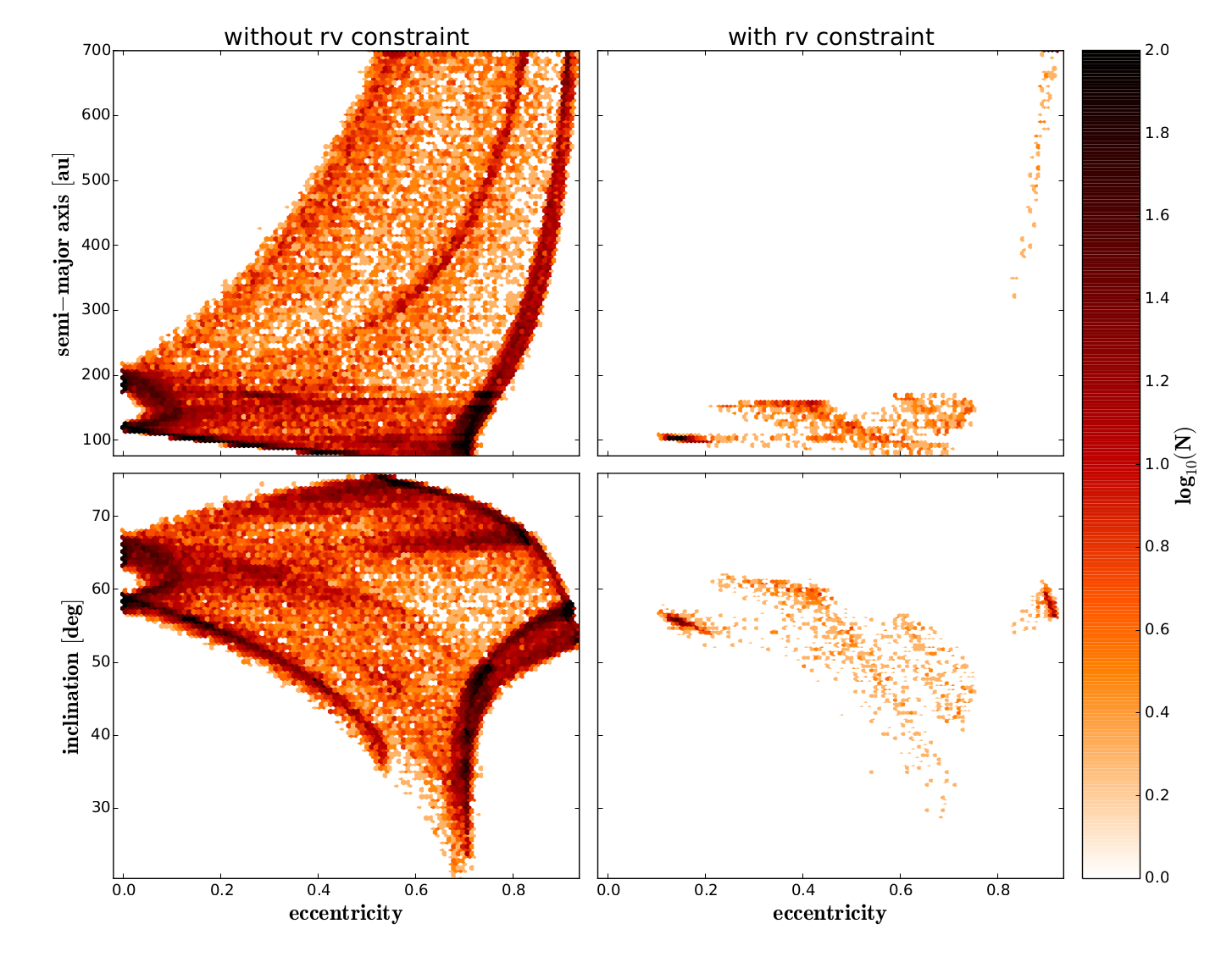}
	\caption{Orbital parameters as function of eccentricity for the sub-stellar companion GQ Lupi b around its host star, with the colour scale indicating the logarithmic density of solutions. The panels in the left column are reproduced from \citet{Ginski2014B}, and they show the $1\%$ best-fitting solutions out of \num{5000000} runs of their LSMC fit. The right-hand panels show the orbital solutions which are consistent with the radial velocity measurement from this work.}  
	\label{fig:orbital-constraints}
\end{figure*}

\subsection{Orbital constraints for GQ Lupi b}
\label{subsec:Orbital-constraints}

The radial velocity measurement of \SI{2.0 \pm 0.4}{\km\per\s} can be used to constrain the orbit of GQ Lupi b. We have approached this by applying the radial velocity as a constraint to the best-fitting orbital solutions from \citet{Ginski2014B}. They applied astrometry to 15 astrometric epochs from VLT/NACO and HST spanning a time frame of 18 years to determine the angular separation and relative position angle. They applied least-squares Monte Carlo (LSMC) statistics to constrain the orbit. The details of the LSMC approach along with a comparison to the more widely used MCMC approach are described in \citet{Ginski2013}. The mass of the host star was assumed to be $0.7 M_{\odot}$, and the distance to the system \SI{140}{pc}. The left-hand panels of Fig. \ref{fig:orbital-constraints} are a reproduction of \citet{Ginski2014B} Fig. 5a and 5b.  The orbital solutions in this figure represent the $1\%$ best-fitting solutions out of \num{5000000} LSMC runs. Dismissing the solutions from this best-fitting subset that are inconsistent with the new radial velocity measurement produces the right-hand panels of Fig. \ref{fig:orbital-constraints}. 

The new radial velocity constraint dramatically reduces the number of possible solutions, and certain families of orbital solutions are excluded. In particular, the observations no longer support:

\begin{enumerate}\itemsep1.2pt
	\item Long-period orbits (a$>$\SI{185}{\au}) with eccentricities below 0.8
	\item Circular orbits
\end{enumerate}

\vspace{0.1 cm}
\noindent The remaining allowed solutions fall into three different families: 

\begin{enumerate}\itemsep1.2pt

  \item Orbits with a semi-major axis $\sim$\SI{100}{\au}, inclination $\sim$\ang{57}, and eccentricity $\sim0.15$

   \item Orbits with a semi-major axis $<$\SI{185}{\au}, a range of eccentricities \num{0.2}$< e <$\num{0.75} and inclinations \ang{28}$< i <$\ang{63}.

   \item Highly eccentric, long-period orbits with a semi-major axis larger than \SI{300}{au}, eccentricities above \num{0.8}, and high inclinations \ang{52}$< i <$\ang{63}.

\end{enumerate}

%%%%%%%%%%%%%%%%%%%%%%
\section{Discussion}
\label{sec:Discussion}
%%%%%%%%%%%%%%%%%%%%%%

\subsection{The slow spin of GQ Lupi b}
\label{subsec:The-slow-spin}

\begin{figure}[ht]
\resizebox{\hsize}{!}{\includegraphics{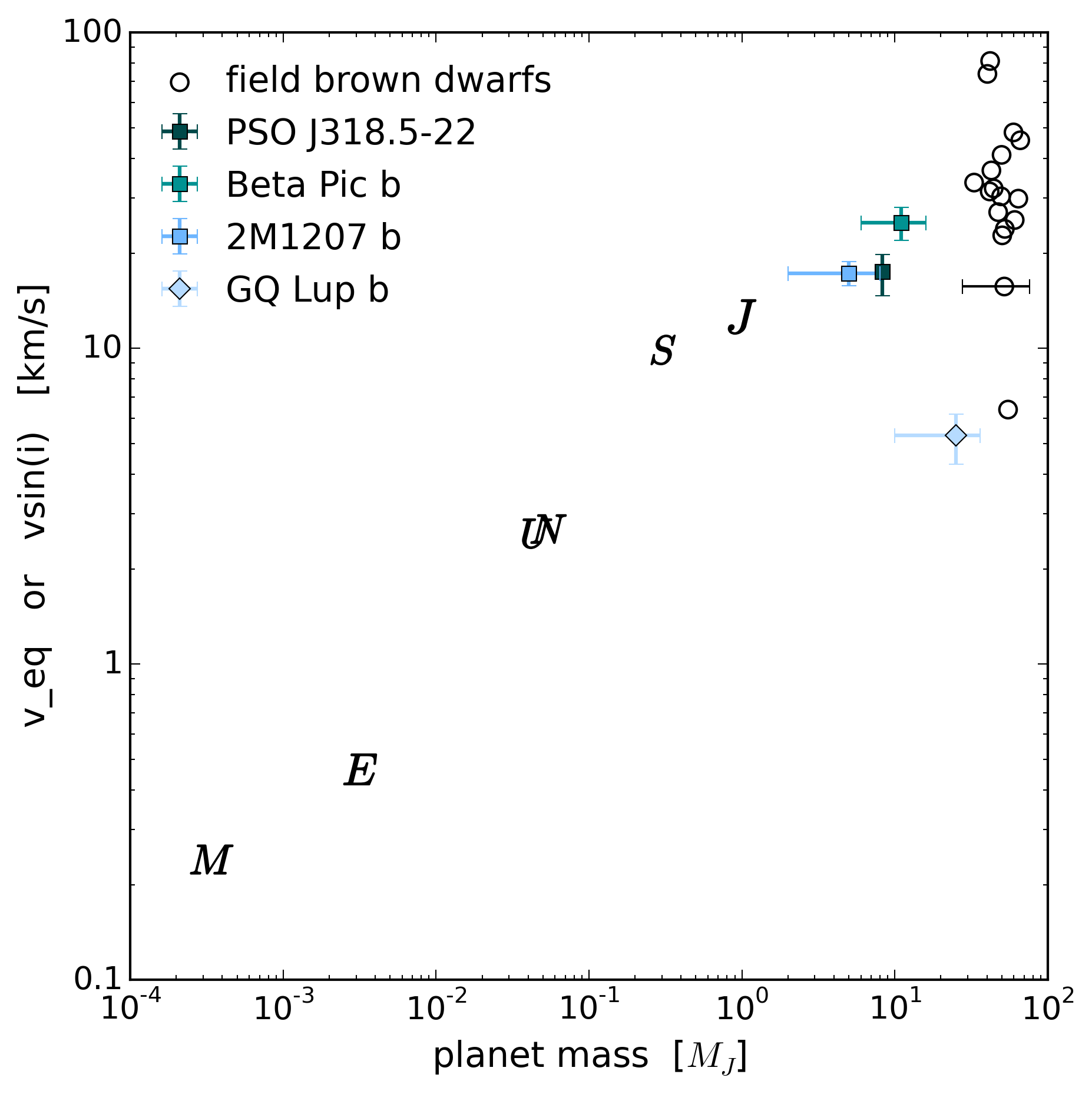}}
	\caption{Spin as function of mass for extrasolar sub-stellar companions and Solar System planets, together with the free-floating planetary mass object PSO J318.5-22 and field brown dwarfs. The Solar System planets and 2M1207 b have equatorial rotation velocities and $\beta$ Pictoris b and GQ Lupi b have projected rotation velocities, as does PSO J318.5-22. Mercury and Venus are not included in the plot because their proximity to the Sun has caused their spin to be dominated by tidal interactions with the Sun. For comparison, field brown dwarfs with estimated masses and either a rotational period measurement or $v\sin(i)$ measurement are shown as empty circles, with the typical uncertainty in mass indicated for a single object. Their masses and radii are from \citet{Filippazzo2015}, rotational periods are from \citet{Metchev2015}, and $v\sin(i)$'s are from either \citet{ZapateroOsorio2006} or \citet{Blake2010}. Note that the field brown dwarfs are much older than the substellar companions and PSO J318.5-22 with highly uncertain ages in the range 500 Myr to 10 Gyr.}
	\label{fig:planet_spin}
\end{figure}

The \ce{CO} and \ce{H2O} lines in the spectrum of GQ Lupi b are narrow, with only a moderate rotational broadening corresponding to a projected rotational velocity of $5.3^{+0.9}_{-1.0}$ \si{\km\per\s}. This strongly suggests that GQ Lupi b is a slow rotator when compared to the giant planets in the Solar System or the recent spin measurements of $\beta$ Pictoris b and 2M1207 b (see Table \ref{table:spin} and Fig. \ref{fig:planet_spin}). 

The $v\sin(i)$ measured for GQ Lupi b could have been influenced by the removal of the stellar lines. E.g., by removing too much of the CO, which actually originates from the companion and not from the star, this removal could have over-subtracted the short-wavelength wing of the companion signal - making it more narrow. We compared the widths of the \ce{CO}+\ce{H2O} signal for the analyses with and without rescaling the stellar CO and found $5.3^{+0.9}_{-1.0}$ \si{\km\per\s} and \SI{4.8 \pm 1.0}{\km\per\s}, respectively. This suggests that the influence in any case is only minor, and cannot have resulted in the companion's $v\sin(i)$ to be artificially small. The measured companion CCFs for the two cases are plotted together in Fig. \ref{fig:CO-adjustment}.

Micro- and macro-turbulence act as additional broadening of line profiles and can therefore cause an overestimation of the projected rotational velocity. \citet{Wende2009} investigated the effective temperature and $\log g$ dependence of the velocity field for M-stars extending to temperatures as low as \SI{2500}{\K}, thereby just covering the estimated temperature of GQ Lupi b. Micro- and macro-turbulent velocities decrease with lower effective temperatures, and particularly micro-turbulence increases towards lower surface gravities. Assuming a surface gravity for GQ Lupi b in the range $\log g =$ \numrange{3.0}{4.2}, a realistic upper limit for the sum of the squares of the turbulent velocities is \SI{1.5}{\km\per\s}, corresponding to an overestimate of $v\sin(i)$ by \SI{0.22}{\km\per\s}. We have included this in the lower bound of the uncertainty estimate of the $v\sin(i)$ measurement.

If the orientation of the spin axis is edge-on, the measured spin velocity corresponds to a rotational period of \SI{82}{hr}, assuming a companion radius of $3.5^{+1.5}_{-1.03}$\si{\Rjup} \citep{Seifahrt2007}. Since the spin axis orientation is unknown, it could possibly be nearly pole on. Therefore the equatorial rotation velocity ($v_{\textrm{eq}}$) could be much higher than $v\sin(i)$. However, the inclination of the axis of rotation would have to be less than \ang{15} to make the spin of GQ Lupi b comparable to $\beta$ Pic b or the Solar System giant planets. Assuming a random orientation, this has only a probability of $3.4\%$.  If we make the assumption that orbital and spin axes are aligned, then based on Fig. \ref{fig:orbital-constraints} we can rule out inclinations below \ang{28}, corresponding to $v_{\textrm{eq}}$ faster than \SI{11.3}{\km\per\s}. This is still well below the $v\sin(i)$ of \SI{25}{\km\per\s}, which \citet{Snellen2014} measured for $\beta$ Pic b. Therefore, it is unlikely that GQ Lupi b is a fast rotator. 

The slow rotation of GQ Lupi b may be caused by a different formation path from $\beta$ Pic b, 2M1207 b, and the giant planets in the solar system. GQ Lupi b at $\sim$\SI{100}{\au} is significantly further away from its host star than the other mentioned planets. In addition it is likely more massive with a mass range \SIrange{10}{36}{\Mjup} extending well into the brown dwarf regime. It could therefore have formed through either disk gravitational instability or even fragmentation of the collapsing proto-stellar core, rather than through core accretion, which is a possible common formation path for the (exo)planets showing the spin-mass trend, we see in Fig. \ref{fig:planet_spin}. \citet{Allers2016} have recently measured the projected rotational velocity of PSO J318.5338-22.8603, which is a free-floating planetary mass member of the $\beta$ Pictoris moving group with an estimated age of \SI{23 \pm 3}{Myr}. Their measured $v\sin(i)=17.5^{+2.3}_{-2.8}$\si{\km\per\s} is also consistent with the Solar System spin-mass trend. Such an object is  most likely the result of gravitational instability \citep{Chabrier2014, Stamatellos2014}, possibly followed by photo-erosion, but it could also have been dynamically kicked out of an orbit around a star, and core accretion can therefore not be ruled out. Low resolution spectra of GQ Lupi b are consistent with a spectral type of late M to early L \citep{Neuhauser2005}, and comparable brown dwarf binaries and field brown dwarfs from the literature show a wide variety of rotational velocities and no simple correlation with mass \citep{Konopacky2012, ZapateroOsorio2006, Metchev2015, Scholz2015}. Most brown dwarfs are rapid rotators ($v\sin(i)>$\SI{10}{\km\per\s}), and their minimum rotation rates are a function of their spectral types, with the higher mass objects rotating more slowly than their lower mass counterparts, in contrast with the planetary trend. This is because magnetic breaking plays an increasing role in the more massive brown dwarfs. However, it is much too early to make any claims about relations between spin, mass and orbital distance based on the limited observations - in particular because there is another very probable explanation for the slow spin of GQ Lupi b.

\begin{table*}[t]
\begin{threeparttable}
\caption{
Comparison of key parameters of the three sub-stellar companions with spin measurements.
}
\label{table:spin}
\centering
\begin{tabular}{l c c c c c c r}
\hline\hline \\
	      & proj. dist. & mass & radius & age & Teff & $v\sin(i)$ or $v_{\textrm{eq}}$ & ref \\
\hline \\
GQ Lup b      & \SI{100}{\au} at \SI{140}{\pc} & $25^{+11}_{-15}$ \si{\Mjup} & $3.5^{+1.5}_{-1.03}$ \si{\Rjup} & $< 5$ \si{Myr} & \SI{2650 \pm 100}{\K} & $5.3^{+0.9}_{-1.0}$ \si{\km\per\s} & 1,2,3 \\
2M1207 b      & \SI{46}{\au} at \SI{59}{\pc} & \SI{5 \pm 3}{\Mjup} & - & \SI{8}{Myr} & \SI{1230 \pm 310}{\K} & \SI{17.3 \pm 1.5}{\km\per\s}$\star$ & 4,5 \\
$\beta$ Pic b & $6-9$ \si{\au} at \SI{19}{\pc} & \SI{11 \pm 5}{\Mjup} & \SI{1.65 \pm 0.06}{\Rjup}      & \SI{21 \pm 4}{Myr} & $1600^{+50}_{-25}$ \si{\K} & \SI{25 \pm 3}{\km\per\s} & 6,7,8,9 \\
\hline
\end{tabular}
\begin{tablenotes}
\small
  \item 1) \citet{Ginski2014B}, 2) \citet{Seifahrt2007}, 3) This work, 4) \citet{Song2006}, 5) \citet{Zhou2016} 6) \citet{Lecavelier2016}, 7) \citet{Binks2014}, 8) \citet{Currie2013}, 9) \citet{Snellen2014}. $\star$ The spin of 2M1207 b is the equatorial rotation velocity based on $P_{\textrm{rot}}=10.7^{+1.2}_{-0.8}$\si{hr} from \citet{Zhou2016} and assuming $R=$\SI{1.5}{\Rjup} consistent with evolutionary models.
\end{tablenotes}
\end{threeparttable}
\end{table*}

We believe that the slow spin of GQ Lupi b is linked to its young age. All current estimates set the age of the GQ Lupi system to $<$\SI{5}{Myr}, and \citet{Neuhauser2005} estimated the age to be only \SI{1 \pm 1}{Myr}. In fact, observational evidence exists of GQ Lupi b actively accreting, through the detection of Pa$\beta$ emission \citep{Seifahrt2007}, as well as the detection of H$\alpha$ in emission together with excess optical continuum emission \citep{Zhou2014}. This means that a significant amount of angular momentum could still be accreted in the future. Furthermore, its radius is now estimated at $3.5^{+1.5}_{-1.03}$\si{\Rjup} \citep{Seifahrt2007}, meaning that even without extra accretion it is expected to still significantly contract to $\sim 1$\si{\Rjup} radius and spin up to $10-25$\si{\km\per\s}. This would bring it much nearer the mass-spin relation of the other giant planets. In this context it is interesting that 2M1207 b with an age in between that of GQ Lupi b and $\beta$ Pic b is also seen to have an intermediate spin. 

The spin angular momenta of brown dwarfs and low-mass stars also evolve over time \citep{Bouvier2014, Irwin2011, Newton2016, Scholz2015}. Analogously to the spin evolution of giant planets, gravitational contraction causes the brown dwarfs to spin up, but after \SIrange{10}{100}{Myr} magnetic breaking starts dominating the spin evolution. The angular momentum loss mechanisms are more inefficient for smaller and cooler objects \citep{Reiners2008}. This stands in contrast to planets, which are expected to retain their spin angular momenta, unless gravitationally disturbed by a third object.

Measurements of spin as a function of mass, orbital distance, and crucially, age, for numerous sub-stellar companions and free-floating low-mass objects, are necessary to determine the effects of formation pathways on spin-mass relationships. Yet at this early stage, every spin measurement of a sub-stellar companion is likely to lead to new insights.

\subsection{The orbital orientation of GQ Lupi b}
\label{subsec:The-orbital-orientation}

We used the radial velocity measurement for GQ Lupi b of \SI{2.0 \pm 0.4}{\km\per\s} in conjunction with previous astrometry from \citet{Ginski2014B} to constrain its orbital elements. This allows us to rule out both circular orbits and long-period orbits ($a>\SI{185}{\au}$) with eccentricities below 0.8. We note that if we allow for a larger uncertainty in the astrometry, then circular orbits with inclinations $\ang{60}<i<\ang{78}$ are still possible. The orbital solutions that are consistent with the RV measurement show a degeneracy between eccentricity and inclination. Orbits with a relatively low eccentricity ($0.1<e<0.4$) have high inclinations $\ang{48}<i<\ang{63}$, which are distinctly different from that of the circumstellar disk ($\sim$\ang{22}). In the case of the more eccentric solutions $0.4<e<0.75$ the inclinations are less well constrained, yet there is a tendency towards allowing for smaller inclinations towards higher eccentricities. This includes orbits with $0.65<e<0.7$ that are near-aligned with the circumstellar disk. Finally we have a subset of orbital solutions with extremely wide orbits (a>\SI{300}{au}) and high eccentricities $e>0.8$. This last subset seems less likely because it would imply that we observe the companion at a special moment near periastron. We note that all displayed solutions in Fig. \ref{fig:orbital-constraints} fit both our new radial velocity measurement and the available astrometric data well, and that the density of solutions does not correspond directly to the likelihood of solutions.

\begin{figure}[ht]
\resizebox{\hsize}{!}{\includegraphics{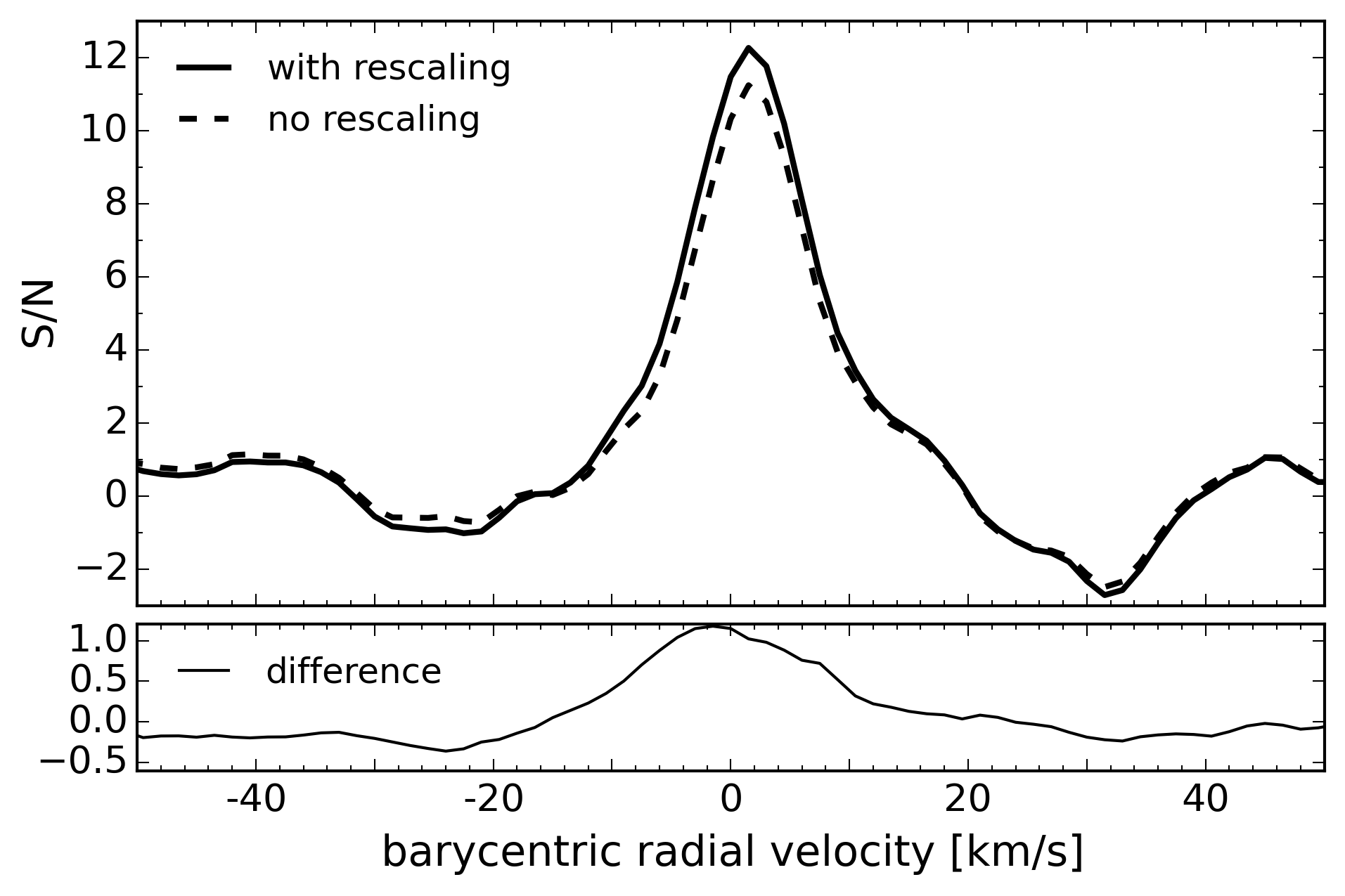}}
	\caption{The measured cross-correlation function between the companion spectrum and the \ce{CO} + \ce{H2O} model with and without the rescaling of the stellar lines described in section \ref{subsec:Companion-position}. The difference between the two stems from when the reference spectrum containing the stellar and telluric lines was removed from the companion spectrum. The solid line is the resulting CCF from the approach we have taken in this paper, where we have scaled the stellar lines in the reference spectrum down by $15\%$ to match the companion to star flux ratio at the companion position. The dashed line is the CCF for the case where the reference spectrum is removed from the companion spectrum without rescaling the stellar lines. The bottom panel displays the difference between the two.}
	\label{fig:CO-adjustment}
\end{figure}

As in the case of the measured $v\sin(i)$, also the RV measurement of the companion can have been influenced by the contribution of the stellar lines. If we have over-subtracted the short-wavelength wing of the companion \ce{CO} lines when we removed the stellar spectrum, this could introduce an artificial redshift. Comparison of the RV measurement with and without the scaling of the stellar contribution (Fig. \ref{fig:CO-adjustment}) indicates that this effect is confined to \SI{0.2}{\km\per\s}. Assuming that we have performed the stellar removal to a precision of $\sim 10\%$, this should not add to the uncertainty in RV to a level of more than \SI{0.1}{\km\per\s}. 

Our RV measurement improves the orbital constraints significantly. A large family of possible orbital solutions found by \citet{Ginski2014B}, in particular with intermediate eccentricities and large semi-major axes, are now no longer in line with the observations. Further constraints from either astrometry or high-dispersion spectroscopy will take at least another decade such that GQ Lupi b has moved significantly enough in its orbit and/or changed its radial velocity. 

The orbital solutions which are misaligned with the circumstellar disk are particularly intriguing. A companion in such an orbit must either have been scattered to that position, or it must have formed in situ, yet outside of the protoplanetary disk. The latter would indicate a formation scenario analogous to that of a binary star system, with the collapsing proto-stellar core fragmenting into smaller objects. 

The orbit of GQ Lupi b is wide enough that disk instability is a possible formation mechanism. Fragmentation of the disk is allowed from approximately \SIrange{50}{300}{\au} \citep{Vorobyov2012}, and although only a small number of fragments are expected to survive to become orbiting planets or brown dwarfs, this is in line with the small number of observed systems that fit the description. HR 8799 and HIP 78530 are other examples. However, if GQ Lupi b formed in situ through disk instability, one would expect low eccentricity orbits within or near the plane of the circumstellar disk. This type of orbit is not supported by previous astrometric analysis by \citet{Ginski2014B}, and our RV measurement strenthens this conclusion. Instead GQ Lupi b may have a high eccentricity orbit which could be an indication that the companion has been scattered to its current position.

\subsection{The systemic velocity and $v\sin(i)$ of GQ Lupi A}
\label{subsec:discussion-host-star-vsini-and-vsys}

We have measured the systemic velocity from the observed spectrum of the star, and find it to be $v_{\textrm{sys}}=\SI{-2.8 \pm 0.2}{\km\per\s}$. The major contributor to the uncertainty is the accuracy of the wavelength solution. \citet{Donati2012} measured the radial velocity in July 2009 to be \SI{-3.2 \pm 0.1}{\km\per\s} and again in June 2011 to be \SI{-2.8 \pm 0.1}{\km\per\s}. They argue that the \SI{0.4}{\km\per\s} change is real and explain it as either long-term changes of the surface granulation pattern or as the result of a third body, a brown dwarf of a few tens of Jupiter masses orbiting the central star at a few au's.

We have measured the $vsin(i)$ of GQ Lupi A to be $\SI{6.8 \pm 0.5}{\km\per\s}$, although without including additional potential broadening mechanisms, which in the case of stars with relatively low projected rotational velocities may contribute significantly to the broadening. The \SI{6.8}{\km\per\s} is therefore likely to be an overestimate. Our result is in agreement with a previous estimate by \citet{Guenther2005} of \SI{6.8 \pm 0.4}{\km\per\s}. \citet{Donati2012} estimated $v\sin(i)$ to be \SI{5 \pm 1}{\km\per\s}, and highlighted that they included magnetic broadening, unlike \citet{Guenther2005}. Also micro-turbulence \citep{Donati2012} and macro-turbulence from sub-surface convective zones \citep{Grassitelli2015} may contribute to the broadening of stellar spectral lines. The rotation period of GQ Lupi A of $8.4$ days \citep{Broeg2007, Donati2012} is typical of classical T Tauri stars, and is often attributed to a disk-locking mechanism \citep{Choi1996, Landin2016}.

%%%%%%%%%%%%%%%%%%%%%
\section{Summary and conclusions}
\label{sec:Conclusion}

The young GQ Lupi system has a central Classical T Tauri star surrounded by a warm dust disk, and is orbited by a sub-stellar companion GQ Lupi b at $\sim$\SI{100}{\au}, which is either a gas giant or a brown dwarf. We observed the parent star and the companion simultanously in the K-band by careful positioning of the slit of the high-dispersion spectrograph CRIRES, in combination with adaptive optics. We made use of both the spatial and spectral information to separate the spectrum of the companion from that of the host star. We detect both water and CO in the companion spectrum. The molecular lines are rotationally broadened and Doppler shifted due to the orbital motion of the companion. We have measured the projected rotation velocity to be $v\sin(i) = 5.3^{+0.9}_{-1.0} $\si{\km\per\s} and the barycentric radial velocity to be RV$=$\SI{2.0 \pm 0.4}{\km\per\s}. 

GQ Lupi b is a slow rotator compared to the giant planets in the Solar System, and to the recent spin measurements of the exoplanets $\beta$ Pic b ($v\sin(i)=$\SI{25}{\km\per\s}, \citet{Snellen2014}) and 2M1207 b  ($v\sin(i)=$\SI{17}{\km\per\s}, \citet{Zhou2016}). This is in spite of GQ Lupi b being likely more massive than either of these, and thus this new spin measurement does not agree with the spin-mass trend of the others. However, we argue that the slow spin is a manifestation of the young age of GQ Lupi, and that the discrepancy cannot be used to argue for fundamental differences in formation path at this time. 

We have used the barycentric radial velocity measurement to place new constraints on orbital elements such as the semi-major axis, the eccentricity and the orbital inclination with respect to the observer. This shows the strength of high-dispersion spectroscopy, because it is possible to measure even small radial velocities. 

The spin and RV measurements of GQ Lupi b demonstrate the potential of the combination of spatial and spectral filtering through the use of high dispersion spectrographs together with adaptive optics.

\begin{acknowledgements}
We are thankful to the ESO staff of Paranal Observatory for their help in performing the observations, and we thank the anonymous referee for the insightful comments and suggestions. This work is part of the research programmes PEPSci and VICI 639.043.107, which are financed by The Netherlands Organisation for Scientific Research (NWO).
Support for this work was provided by NASA through Hubble Fellowship grant HST-HF2-51336 awarded by the Space Telescope Science Institute, which is operated by the Association of Universities for Research in Astronomy, Inc., for NASA, under contract
NAS5-26555. Furthermore, this work was performed in part under contract with the Jet Propulsion Laboratory (JPL) funded by NASA through the Sagan Fellowship Program executed by the NASA Exoplanet Science Institute. 
\end{acknowledgements}

%%%%%%%%%%%%%%%%%%%%%%%%%% BIBLIOGRAPHY %%%%%%%%%%%%%%%%%%%%%%%%%%%%%%%%%%%%%%%%%%
% Include full biography list
%\nocite{*}
% Textual citation e.g. Jones et al. (1990)
%\citet{jon90}
% Parenthetical citation e.g. (Jones et al. 1990)
%\citep{jon90}

\bibliographystyle{aa}
\bibliography{mybib}

\end{document}